\documentclass[11pt,a4paper]{article}
\usepackage{jheppub}

\usepackage{hyperref}

\title{Logarithms and Volumes of Polytopes}
\author[a]{Michael Enciso}

\affiliation[a]{Mani L. Bhaumik Institute for Theoretical Physics\\
Department of Physics and Astronomy\\
University of California at Los Angeles\\
Los Angeles, CA 90095, USA }

\emailAdd{menciso@physics.ucla.edu}

\abstract{ Describing the geometry of the dual amplituhedron without reference to a particular triangulation is an open problem.  In this note we introduce a new way of determining the volume of the
   tree-level NMHV dual amplituhedron.  We show that certain contour integrals of logarithms serve as natural building blocks for computing this volume as well as the volumes of general polytopes in any dimension.  These building blocks encode the geometry of the underlying polytopes in a triangulation-independent way, and make identities between different representations of the amplitudes manifest.
  }   
\begin{document}
\maketitle

\section{Introduction}

Recent years have seen tremendous progress in
understanding scattering amplitudes in both gauge and gravity
theories.  New mathematical structures that are not apparent in
textbook formulations of quantum field theory have been uncovered, and
many computations have been immensely streamlined in comparison with
the standard Feynman diagram approach (see the recent reviews~\cite{ElvangRev,DixonRev, HennRev} and
references therein).  While many of these developments have
applications in theories with various amounts of (including no)
supersymmetry, the computational simplicity of maximally
supersymmetric gauge and gravity theories make them ideal testing
grounds for new ideas~\cite{SimplestQFT}.

One of the major breakthroughs in the study of maximally
supersymmetric gauge theories is the discovery of the amplituhedron,
an object that encodes all tree-level amplitudes and loop-level
integrands in planar $\mathcal{N} = 4$ super-Yang--Mills theory
(sYM)~\cite{Amplituhedron,IntoAmplituhedron}.  Schematically, and specializing to the case of
tree amplitudes, the amplituhedron is a region of a particular
positive Grassmannian~\cite{Amplituhedron, BigGrassmannian,Postnikov}.  This region encodes the amplitude
via a volume form with logarithmic singularities on its boundary, and
after stripping off a canonical prefactor from this form what remains
(up to some fermionic integrations) is the amplitude.  For loop
integrands the same is true but with the amplituhedron corresponding
to a region of a particular generalization of the positive
Grassmannian.  In the rest of this note we restrict ourselves to the
tree-level case.

For tree-level NMHV amplitudes, the amplitude obtained in this way
is naturally interpreted as the volume of a polytope in a $\mathbb{CP}^4$ that is dual to the space in which the amplituhedron lives~\cite{NoteOnPolytopes,Amplituhedron}. N$^k$MHV
tree amplitudes with $k\geq1$ are therefore viewed as a type of ``generalized volume'' of a dual amplituhedron
\cite{NoteOnPolytopes,PositiveAmplituhedron}.  For $k>1$ a geometric understanding of the dual
amplituhedron is unclear, though there are strong indications that
such a picture should exist \cite{PositiveAmplituhedron, TowardsVolume}.

In this note we introduce a new way of computing the volume of the
tree-level NMHV (or $k=1$) dual amplituhedron directly in the space in which the polytope lives.  The basic objects in this method are contour integrals with simple, closed
contours in the complex projective space containing the polytope.  In Ref. \cite{NoteOnPolytopes} the authors computed these volumes by integrating a particular volume form over the underlying polytope in
the dual space, thus placing the information about the polytope in the
contour (which has boundaries).  As we will see in section 3, our method differs from that in Ref. \cite{NoteOnPolytopes} by using contours that are
closed (i.e., without boundary) and canonically specified by the integrands themselves.  This is in contrast to, for example, ``dlog'' representations of amplitudes, where the contour is not specified by the integrand itself~\cite{BigGrassmannian}.  Additionally, the method we introduce is independent of any particular triangulation of the underlying polytope, and can be used to recover any such triangulation.

In Ref.~\cite{OldPaper} we provided a definition of ``combinatorial
polytopes'' which incorporates a general class of polytopes. For
these polytopes neither convexity (and therefore positivity) nor even
connectivity are necessary.  We introduced a set of new objects that we denote by $F_{i_1...i_n}$ and will now refer to as ``vertex objects.''  The reason for this naming convention is that the subscripts of these vertex objects correspond to the vertices of polytopes in a natural way that we will review shortly.  In Ref.~\cite{OldPaper} we showed that we obtain the volume of a polytope by summing these vertex objects over the vertices of the polytope.  This way of expressing
the volume of a polytope does not require any triangulation of the
polytope to be known, and the volume of the polytope is uniquely
expressed in terms of these vertex objects.  These observations
motivate us to view the vertex objects as basic building blocks for
computing volumes of polytopes.

The vertex objects satisfy a simple relation that allows us to easily derive many non-trivial identities between different representations of the tree-level NMHV amplitude, as we will review in the next section.  These identities and their more complex analogues for N$^k$MHV amplitudes with $k>1$ can also be
derived using global residue theorems (GRTs) on an auxiliary Grassmannian \cite{MasonSkinner,DualityForSMatrix}.    In this picture, computing tree amplitudes and loop integrands is equivalent to specifying the correct contour for a particular integrand in the Grassmannian~\cite{BigGrassmannian}, and relations between different representations of the amplitude follow from the GRTs. Introducing this auxiliary space manifests the Yangian symmetry of the amplitudes~\cite{HennPlefka}, while the geometry of the underlying space whose volume corresponds to the amplitude gets obscured.  By showing that the vertex objects discussed above are naturally given by contour integrals in the dual space directly, we give a formalism that both manifests the relations between different representations of the amplitude while avoiding the introduction of an auxiliary space.  This formalism has not been extended to N$^k$MHV amplitudes with $k>1,$ but doing so will likely illuminate the underlying geometry of the dual amplituhedron.

The outline of this note is as follows: In the next section we briefly
review some key properties of complex projective space and the
standard generalization of volumes of polytopes to projective spaces.
We will also briefly describe how NMHV tree amplitudes are expressed
as volumes of polytopes and how the vertex objects are defined and
used.  In section 3 we show how contour integrals of logarithms
naturally arise in computing the areas of quadrilaterals and their
higher-dimensional analogues.  In section 4 we show how the vertex
objects correspond to a particular combination of these integrals.

\section{Polytopes in Projective Space}

In this section we review the ideas that will be needed in later sections.  After discussing some key facts about (complex) projective spaces, we will review the standard generalization of volumes of polytopes in affine space to that of polytopes in projective space.  We then briefly describe the formalism introduced in Ref.~\cite{OldPaper}, where the vertex objects encode the geometry of polytopes as well as give their volumes.  Finally, we review how these vertex objects are used to manifest certain properties of the NMHV tree-amplitude.  In the remaining sections of this note we show how these vertex objects are given as contour integrals in the space containing the polytope.

\subsection{Projective Geometry}

In this brief review of projective geometry we follow Ref.~\cite{Huggett} and the first appendix of Ref.~\cite{CP5Calculus}, which provide more complete discussions of these ideas.  

A point $\mathbb{Z}^{\alpha}\in \mathbb{CP}^n$ is defined by $n+1$ homogenous coordinates, one for each value of $\alpha=0, ..., n.$  Each such point defines an $(n-1)$-dimensional hyperplane $H_Z$ in the dual $\mathbb{CP}^{n*}$ by placing a single linear constraint on the homogenous coordinates of the dual elements.   Namely, we have \begin{equation}H_Z\ \equiv \ \{A_{\alpha}\in \mathbb{CP}^{n*}\ |\ Z\cdot A\equiv Z^{\alpha}A_{\alpha}=0\}\simeq \mathbb{CP}^{n-1}\subset \mathbb{CP}^{n*}.\end{equation}  The subspace $H_Z$ is a linearly embedded $\mathbb{CP}^{n-1}$ in the dual $\mathbb{CP}^{n*}.$  We will refer to linearly embedded $\mathbb{CP}^1$'s, $\mathbb{CP}^2$'s, and $\mathbb{CP}^k$'s with $k>2$ respectively as lines, planes, and hyperplanes, even though the underlying topology of these spaces may be rather different.  For example, a $\mathbb{CP}^1$ is a Riemann sphere though we will still refer to it as a line.

\begin{figure}[h!]
\centering
    \includegraphics[scale=.4]{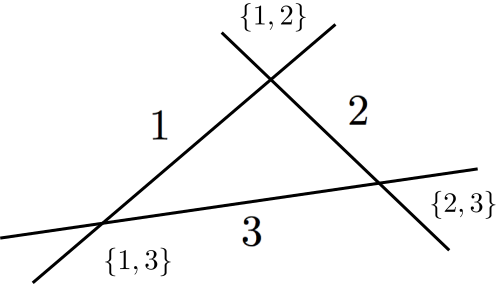}
        \caption{Three lines in $\mathbb{CP}^{2*}$ labeled by $i$ corresponding to three points $\{Z_i^{\alpha}\}_{1\leq i \leq 3}$ in $\mathbb{CP}^2$.  The intersection of lines $i$ and $j$ is denoted by $\{i,j\}.$  We note that $\{i,j\}=\{j,i\}$ implicitly.}\label{TriangleIntersections}
\end{figure}

Intersections of lines, planes, and hyperplanes always exist in projective geometry.  For example, three points $Z_1^{\alpha},$ $Z_2^{\alpha},$ and $Z_3^{\alpha}$ in $\mathbb{CP}^2$ give three lines in the dual $\mathbb{CP}^{2*}$ and each pair of lines intersects in a unique point.  This is shown in Figure \ref{TriangleIntersections}, where the line dual to $Z_i^{\alpha}$ is labeled by $i,$ and the intersection of lines $i$ and $j$ is labeled by $\{i,j\}.$ 

More generally, any two distinct $(n-1)$-dimensional hyperplanes in $\mathbb{CP}^{n*}$ intersect in a unique $(n-2)$-dimensional hyperplane.  Namely, two points $Z_1^{\alpha}$ and $Z_2^{\alpha}$ in $\mathbb{CP}^n$ define two $(n-1)$-dimensional hyperplanes $H_{Z_1}$ and $H_{Z_2}$ in $\mathbb{CP}^{n*},$ and we have that \begin{equation}H_{Z_1}\cap H_{Z_2}\simeq \mathbb{CP}^{n-2}\subset \mathbb{CP}^{n*}. \end{equation}   We therefore see that $n$ distinct points in $\mathbb{CP}^n$ uniquely define a point in the dual $\mathbb{CP}^{n*}$ via the simultaneous intersection of their $n$ dual hyperplanes.

\subsection{Volumes of Simplices}

There is a natural generalization of the volume of a polytope to projective space.  By first understanding this extension for the case of a simplex, the volume of more general polytopes follows immediately by considering sums of simplices.  We will therefore follow Ref.~\cite{NoteOnPolytopes} and review how to express the volume of simplices in a projective way. 

\begin{figure}[h!]
\centering
    \includegraphics[scale=.4]{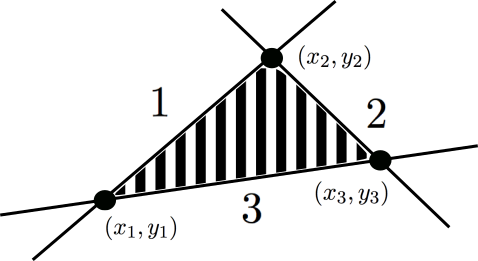}
        \caption{A triangle in affine space defined by vertices $(x_i,y_i),$ with faces (edges) labeled by $i$ corresponding to the points $\{Z_i^{\alpha}\}_{1\leq i\leq 3}$ in the dual space that define them.  The $Z_i^{\alpha}$'s are explicitly defined in terms of the $(x_i,y_i)$ coordinates in the text.}\label{AreaOfTriangle}
\end{figure}

We begin by considering the area of a two-simplex, or a triangle, in real affine space with vertices located at $(x_1,y_1),$ $(x_2,y_2),$ $(x_3,y_3),$ as shown in Figure \ref{AreaOfTriangle}.  We can write its area $A$ as \cite{NoteOnPolytopes} 
\begin{equation}A=\frac{1}{2}\frac{\langle Z_1 Z_2 Z_3\rangle^2}{\langle Z_1 Z_2 P\rangle\langle Z_2 Z_3P\rangle\langle Z_3 Z_1P\rangle}\equiv[123], \label{3Bracket}\end{equation} where we have introduced the notation $\langle Z_1...Z_n\rangle\equiv \varepsilon_{\alpha_1...\alpha_n}Z_1^{\alpha_1}...Z_n^{\alpha_n},$ with the value of $n$ taken from context.  We have also defined \begin{align}    W_{i\alpha}\equiv \begin{pmatrix}x_i\\y_i\\1\end{pmatrix}, \ \ \ \ P^{\alpha}=\begin{pmatrix} 0\\0\\1\end{pmatrix} \end{align} as well as \begin{align} 
Z_1^{\alpha}\equiv  \varepsilon^{\alpha\beta\gamma}W_{1\beta}W_{2\gamma}, \ \ Z_2^{\alpha}\equiv  \varepsilon^{\alpha\beta\gamma}W_{2\beta}W_{3\gamma}, \ \ Z_3^{\alpha}\equiv  \varepsilon^{\alpha\beta\gamma}W_{3\beta}W_{1\gamma}.
\end{align} 
We note that the $Z_i^{\alpha},$ $W_{i\alpha},$ and $P^{\alpha}$ all have three homogenous coordinates, in line with their being elements of $\mathbb{CP}^2$ (or its dual).  We have simply ``lifted'' the affine coordinates into a particular coordinate patch of projective space by placing a 1 in the third component of the $W_{i\alpha}$'s.

Equation (\ref{3Bracket}) is projectively well-defined in the $Z_i^{\alpha}$'s---which, according to the discussion in the previous subsection, determine the faces of the triangle---thus allowing their domain of definition to extend to $\mathbb{CP}^2.$  We note that (\ref{3Bracket}) is not projectively well-defined in $P^{\alpha}$ since it defines the line at infinity in $\mathbb{CP}^{2*}$ and therefore the scaling of the area---the scaling we choose here corresponds to the choice of placing 1 (as opposed to a different non-zero number) in the third component of the $W_{i\alpha}$'s.  Equation (\ref{3Bracket}) is also completely antisymmetric in the $Z_i^{\alpha}$'s, corresponding to the two possible orientations of the triangle.

It will be instructive to see explicitly how this works for
one-dimensional simplices as well.  A one-simplex is simply a line, and the distance $L$ between two
points $x_1$ and $x_2$ in $\mathbb{R}$ can be written
as 
\begin{equation}L=\frac{\langle Z_1 Z_2\rangle}
 {\langle Z_1P\rangle  \langle Z_2 P\rangle}.
\label{finalLengthOfLine}
\end{equation}
Here we have defined \begin{equation} Z_1^{\alpha}\equiv \varepsilon^{\alpha\beta}W_{i\beta}, \ \  \ W_{i \alpha}\equiv \begin{pmatrix}x_i\\ 1 \end{pmatrix}, \ \text{and}\ \  \ P^{\alpha}\equiv \begin{pmatrix}0\\ 1 \end{pmatrix}. \label{1DDefinitions}\end{equation}  Equation (\ref{finalLengthOfLine}) indeed reproduces $L=x_1-x_2,$ as expected, and it expresses the length of the line defined by the endpoints $W_{1\alpha}$ and $W_{2\alpha}$ in terms of their duals and the point at infinity defined by $P^{\alpha}.$  It is projective and antisymmetric in $Z_1^{\alpha}$ and $Z_2^{\alpha},$ corresponding to the two different orientations of the line.

This generalizes to volumes of simplices in any dimension.  For any
$D+1$ points $\{Z_i^{\alpha}\}_{1\leq i\leq D+1}$ in $\mathbb{CP}^D$ there are $D+1$
hyperplanes in the dual $\mathbb{CP}^{D*},$ and the volume of the
simplex bounded by these hyperplanes is given by~\cite{NoteOnPolytopes} \begin{equation}
  V=\frac{1}{D!}\frac{\langle Z_1...Z_{D+1}\rangle^D}{\langle
    Z_{1}...Z_{D}P\rangle\langle Z_{2}...Z_{D+1}P\rangle...\langle
    Z_{D+1}...Z_{D-1}P\rangle}\equiv [12...(D+1)].\end{equation} This
expression is projective and totally antisymmetric in the $Z_i^{\alpha}$'s.  The antisymmetry corresponds to the two possible
orientations of the simplex.

The dimension most relevant for scattering amplitudes is four, so for completeness we will explicitly write the volume of a four-simplex, bounded by the five faces defined by $Z_1^{\alpha},..., Z_5^{\alpha}.$  Translating the above formula gives 
\begin{equation}V=\frac{1}{4!}\frac{\langle Z_1 Z_2 Z_3 Z_4 Z_5 \rangle^4}{\langle Z_1 Z_2 Z_3 Z_4 P \rangle\langle Z_2 Z_3 Z_4 Z_5 P \rangle\langle Z_3 Z_4 Z_5 Z_1 P \rangle\langle Z_4 Z_5 Z_1 Z_2 P \rangle\langle Z_5 Z_1 Z_2 Z_3 P \rangle} \equiv [12345]. \end{equation}

\subsection{Volumes of General Polytopes}\label{GeneralPolys}

\begin{figure}[h!]
\centering
    \includegraphics[scale=.3]{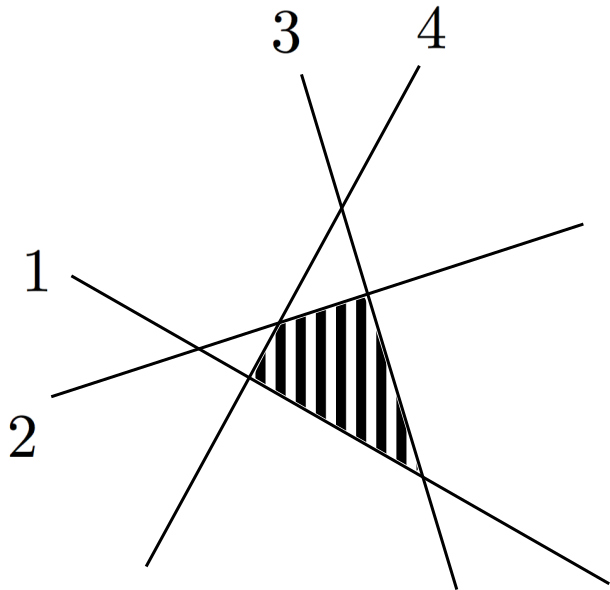}
        \caption{A quadrilateral in $\mathbb{CP}^{2*}$ defined by four lines labeled by $i$ according to the points $\{Z_i^{\alpha}\}_{1\leq i\leq 4}$ in $\mathbb{CP}^2$ that define them.}\label{AreaOfQuadrilateral}
\end{figure}

For a fixed dimension $D,$ we can view any sum of simplices as the volume of a general polytope, expressed through some particular triangulation.  For example, four points $Z_1^{\alpha},$ $Z_2^{\alpha},$ $Z_3^{\alpha},$ and $Z_4^{\alpha}$ in $\mathbb{CP}^2$ define four lines in the dual $\mathbb{CP}^{2*}.$  These four lines are depicted in Figure \ref{AreaOfQuadrilateral} and are respectively labeled by 1, 2, 3, and 4. 

\begin{figure}[h!]
\centering
    \includegraphics[scale=.25]{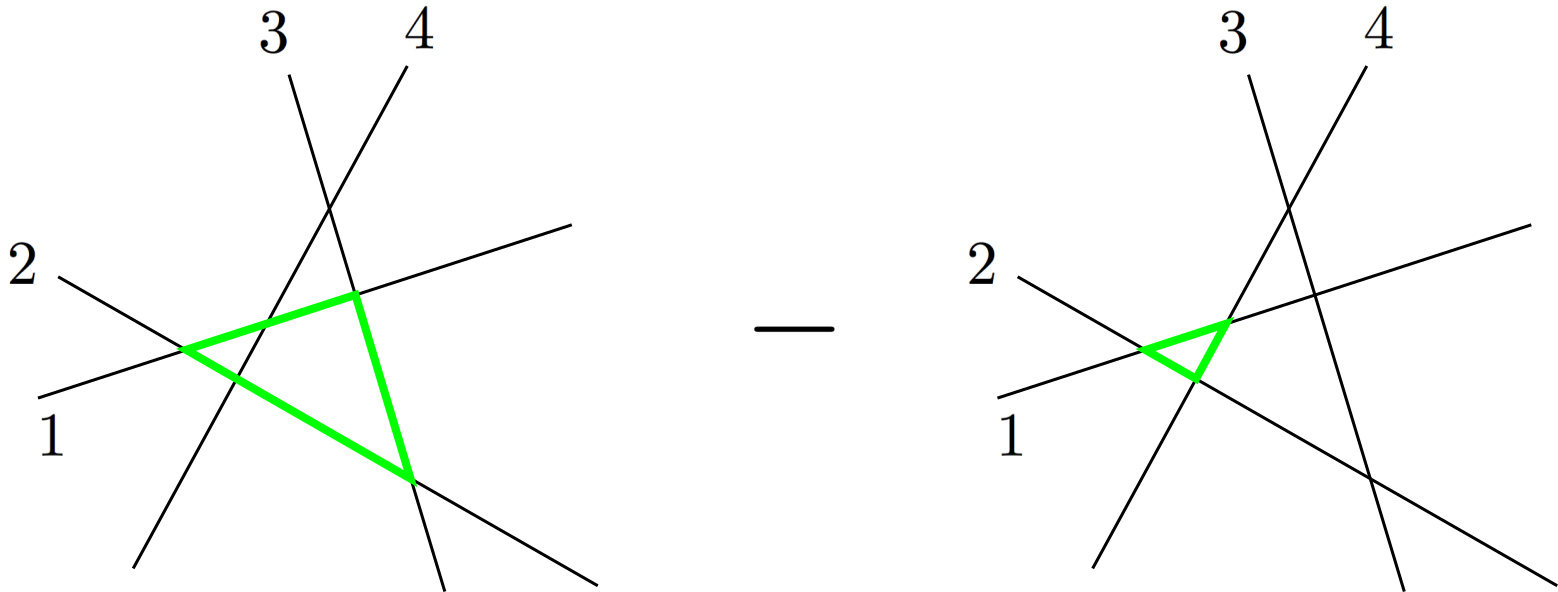}
        \caption{The quadrilateral shown in Figure \ref{AreaOfQuadrilateral} viewed as the difference of two triangles.}\label{DifferenceOfTriangles}
\end{figure}

The area of the shaded quadrilateral can be written as \begin{equation}A=[123]-[124]\label{InformalQuadrilateralArea},\end{equation}which is the area of the triangle bounded by the faces 1, 2, and 3 minus the area of the triangle bounded by the faces 1, 2, and 4.   This is depicted in Figure \ref{DifferenceOfTriangles}.

By inspection of Figure \ref{AreaOfQuadrilateral} we also see that we can write the area of the same region as \begin{equation}A=[431]-[432],\end{equation} by viewing this area as the difference between the area of the triangle bounded by the faces 4, 3, and 1, and the triangle bounded by the faces 4, 3, and 2.  We therefore see that we have \begin{equation}[123]-[124]=[431]-[432], \label{nonTrivialTriangles}\end{equation} which, when one unravels the definition of these 3-brackets, is a non-trivial relation.

Proving this relation through repeated application of Schouten identities on the $\langle...\rangle$ brackets quickly shows that this geometric proof is more convenient, especially for analogous relations in higher dimensions.  However, this geometric proof is not very precise, for a few reasons.  For one, we have not been careful to keep track of the orientation of the quadrilateral in our two different triangulations.  A second and more serious ambiguity is that our notion of a polytope itself is rather tenuous.  Namely, once we extend our underlying space from a real affine space to a complex projective space, any notion of ``inside'' or ``outside'' is lost.  Moreover, one generally thinks of a $D$-dimensional polytope in a $D$-dimensional space as being some full-dimensional region carved out by a finite number of hyperplanes.  However, by complexifying our compact space, we end up talking about $D$-dimensional polytopes in $\mathbb{CP}^{D},$ which is a space of $2D$ real dimensions.  A third issue with trying to define a polytope as a sum of volumes of simplices is that there are (infinitely) many triangulations that correspond to the same polytope.  Some triangulations may make apparent certain geometric qualities of the underlying polytope while masking others.

The amplituhedron makes precise sense of these polytopes as a region in a positive Grassmannian, and for the NMHV case under consideration, this Grassmannian is simply a projective space~\cite{Amplituhedron}.  In this program one considers convex polytopes, which places positivity constraints on the external kinematics.  One then analytically continues to consider general kinematics.  In Ref.~\cite{OldPaper} we instead focused solely on the combinatorial structure of polytopes.  We then gave a precise definition of a general type of polytope that is not necessarily convex or even connected.  In the next subsection we will briefly review these ideas in two dimensions, as well as introduce the two-dimensional vertex objects $\{F_{ij}\}.$  We refer to Ref.~\cite{OldPaper} for details and the higher-dimensional cases.

\subsection{The Vertex Formalism}

\begin{figure}[h!]
\centering
    \includegraphics[scale=.3]{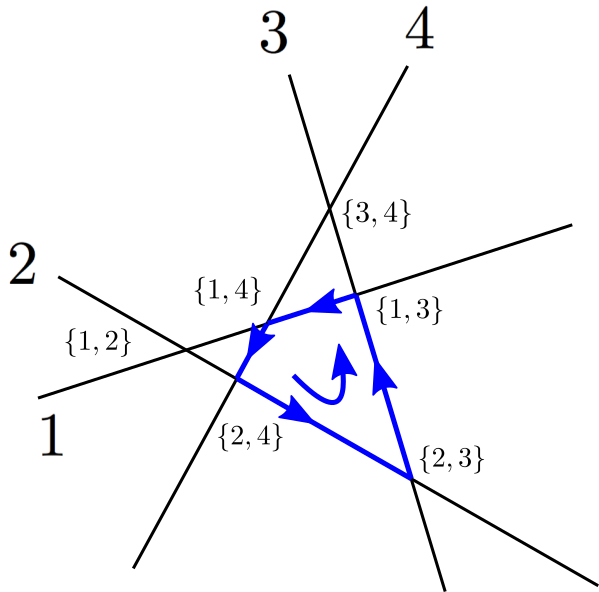}
        \caption{The quadrilateral depicted in Figure \ref{AreaOfQuadrilateral} defined solely through the intersection of its faces.}\label{QuadrilateralList}
\end{figure}

We consider again the quadrilateral in Figure \ref{AreaOfQuadrilateral} and our goal will be to give it a precise definition.  While this figure does not correctly depict the topology of the objects involved---as mentioned above, the lines are actually Riemann spheres---it does correctly depict the intersection structure of these objects.   We therefore define this polytope by its intersection structure, saying that this is the ``quadrilateral'' defined by starting at the vertex $\{1,4\}$ and walking along line 4 to arrive at the vertex $\{2,4\},$ then walking along line 2 to arrive at the vertex $\{2,3\},$ then walking along line 3 to arrive at the vertex $\{3,1\},$ and then walking along line 3 to arrive back at the vertex $\{1,4\}.$  This is depicted in Figure \ref{QuadrilateralList}.

This set of instructions can be succinctly summarized by the list $(1423),$ which we define to be shorthand for \begin{equation}\{1,4\} \rightarrow \{4,2\}\rightarrow \{2,3\}\rightarrow \{3,1\} \rightarrow \{1,4\} \label{VertexList}\end{equation} where each ``$\rightarrow$'' means to travel along the line whose label is common to the vertex on either side of the arrow.

In Ref.~\cite{OldPaper} we introduced a collection $\{F_{ij}\}$ of vertex objects defined as a particular sum of volumes of simplices.  These objects are referred to as vertex objects because a vertex of a two-dimensional polytope is labeled by two lines, as is each $F_{ij}.$  We found that these vertex objects are antisymmetric, so that $F_{ij}=-F_{ji},$ and that they satisfy\footnote{These vertex objects differ from those introduced in Ref.~\cite{OldPaper} by a factor of 2.} 
\begin{equation}
F_{ij}+F_{jk}+F_{ki}=[ijk]\label{Intro2DMonodromy}
\end{equation}
for any choice of $i,$ $j,$ and $k,$ where we recall that $[ijk]$ is the volume of the two-simplex bounded by the three lines $i,$ $j,$ and $k$.

We consider the sum $F_{14}+F_{42}+F_{23}+F_{31}$ over the vertices of this quadrilateral.  Using the antisymmetry of each $F_{ij}$ and equation (\ref{Intro2DMonodromy}), we find\begin{align}F_{14}+F_{42}+F_{23}+F_{31}&=F_{14}+F_{42}+([231]-F_{12}) \label{firstLine} \nonumber \\&=[123]-(F_{12}+F_{24}+F_{41}) \nonumber \\&=[123]-[124], \end{align} which is precisely the volume of the quadrilateral that the list of vertices in (\ref{VertexList}) defines.  Applying equation (\ref{Intro2DMonodromy}) to the left hand side of (\ref{firstLine}) in a different order also shows that \begin{equation}F_{14}+F_{42}+F_{23}+F_{31}=[431]-[432].\end{equation}  This gives a quick and rigorous proof of the non-trivial identity (\ref{nonTrivialTriangles}).  Indeed, all possible triangulations of the quadrilateral can be obtained by applying (\ref{Intro2DMonodromy}) to the left hand side of (\ref{firstLine}), giving a simple algebraic method for proving many non-trivial identities amongst sums of simplices~\cite{OldPaper}. 

\begin{figure}[h!]
\centering
    \includegraphics[scale=.2]{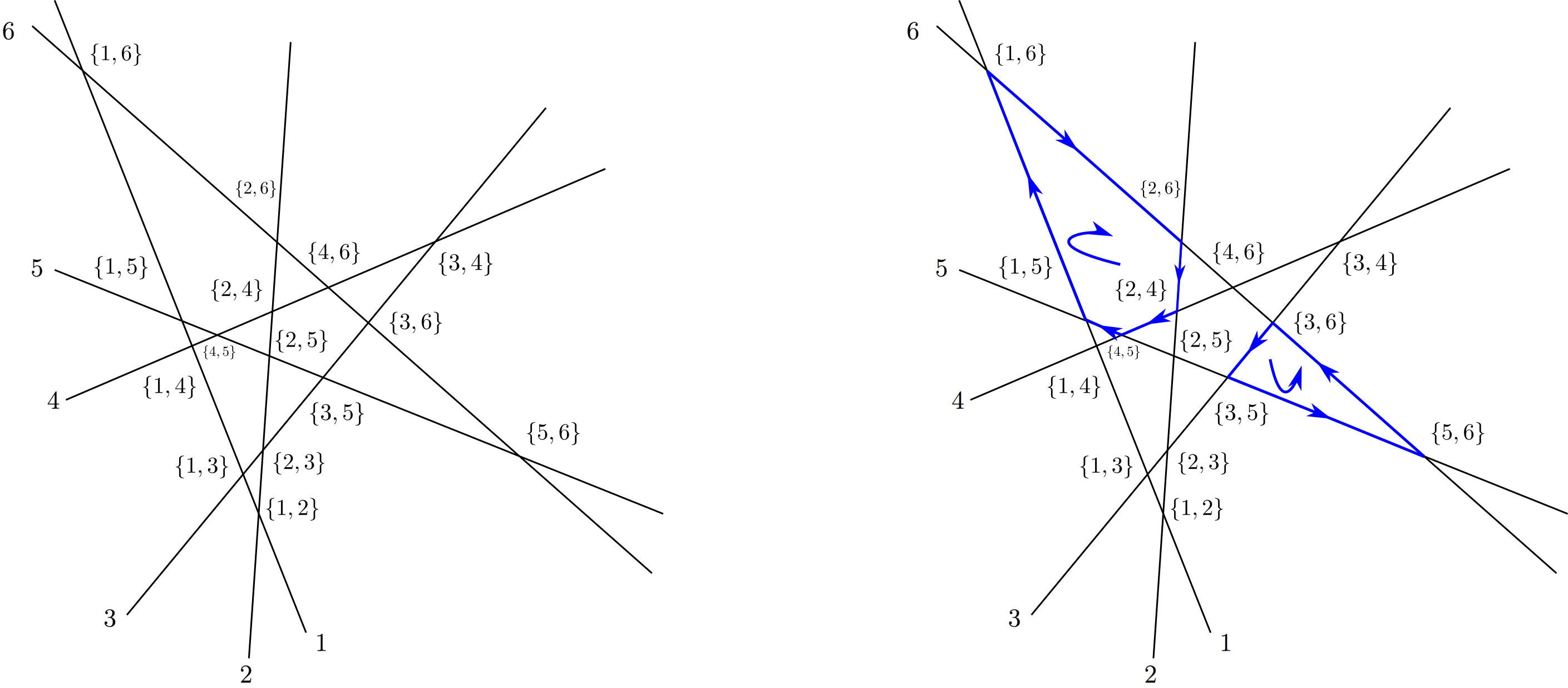}
        \caption{A general polygon (right) defined solely through a set of instructions for traversing the intersections of six lines (left).  Lines are labeled by $i$ according to the $Z_i^{\alpha}$ that defines them, and the intersection of two lines $i$ and $j$ is denoted by $\{i,j\}.$  It is implied that $\{i,j\}=\{j,i\}.$  }\label{GeneralPolygon}
\end{figure}

This example is a special case of a more general phenomenon---given any set of vertex-connecting instructions defining any polygon, summing the corresponding $F_{ij}$ for each vertex yields the area of that polygon.  This process works for general polygons, even disconnected ones. 
 For example, suppose we have six elements $\{Z_{i}^{\alpha}\},$ $1\leq i\leq 6,$ defining six lines, as shown on the left hand side of Figure \ref{GeneralPolygon}.  We can then define the disconnected polygon shown on the right hand side of this figure by the instructions \begin{align} \{5,1\} \rightarrow \{1,6\}\rightarrow \{6,2\}\rightarrow \{2,4\}\rightarrow \{4,5\}\rightarrow \{5,6\}\rightarrow \{6,3\}\rightarrow \{3,5\}\rightarrow \{5,1\}. \end{align}  Analogously to the case of the quadrilateral, this set of instructions corresponds to the list $(51624563).$  It is then the case, rather surprisingly, that the area $A$ of this polygon can be written simply as \begin{equation}A=F_{51}+F_{16}+F_{62}+F_{24}+F_{45}+F_{56}+F_{63}+F_{35}.\end{equation}  This can be checked against any particular triangulation of this polygon.  Additionally, any triangulation of this polygon can be obtained from this expression through repeated use of (\ref{Intro2DMonodromy}).

The general result can be stated as follows.  Let $\{Z_i^{\alpha}\}$ be a collection of $N$ elements in $\mathbb{CP}^2$ defining $N$ lines in the dual $\mathbb{CP}^{2*}.$  The most general polygon in this dual $\mathbb{CP}^{2*}$ is given by a list $(i_1i_2 ... i_n),$ corresponding to the instructions \begin{equation}\{i_1,i_2\} \rightarrow \{i_2,i_3\}\rightarrow ... \rightarrow \{i_{n-1},i_n\}\rightarrow \{i_n,i_1\}\rightarrow \{i_1,i_2\}.\end{equation}  The area $A$ of this polygon is then given by the following sum over the vertices: \begin{equation}A=\sum_{k=1}^nF_{i_ki_{k+1}},\label{AreaOfGeneralPolygon}\end{equation} and from this expression any particular triangulation can be obtained~\cite{OldPaper}.  This form of the area is independent of any particular triangulation and is inherently tied to the data of the polygon itself---its vertices and how we traverse them.  

We note that many different lists give rise to the same polygon.  For example, any cyclic permutation of a list gives the same polygon.  More trivially, the list $(1234)$ is identical to the list $(12121234),$ since the latter corresponds to staying on the vertex $\{1,2\}$ over and over again before moving on.  However, the final result in terms of the vertex objects (up to trivial cancellations) is identical.  For example, the sum of these objects corresponding to the list $(12121234)$ is simply \begin{equation}F_{12}+F_{21}+F_{12}+F_{21}+F_{12}+F_{23}+F_{34}+F_{41},\end{equation} which, after using the antisymmetry of $F_{ij}$ yields the same result as the list $(1234).$  Indeed, the sum in (\ref{AreaOfGeneralPolygon}) is dependent only on the equivalence class of lists, where equivalence of lists is defined by their determining the same polygon.  In Ref.~\cite{OldPaper} we show how to extend this definition of polygon to arbitrary higher-dimensional polytopes.

In Ref.~\cite{OldPaper} we also defined the corresponding vertex objects in higher dimensions.  For example, in three dimensions we defined a collection of vertex objects $\{F_{ijk}\}$ completely antisymmetric in their subscripts and satisfying \begin{equation}F_{ijk}-F_{jkl}+F_{kli}-F_{lij}=[ijkl] \label{Review3DMonodromy}  \end{equation} for any choice of $i,j,k,l.$  We continue to use the term ``vertex objects'' because for a three-dimensional polytope a vertex is defined by the intersection of three planes, each defined by a $Z_i^{\alpha},$ and these planes determine the subscripts of a given $F_{ijk}.$   In four dimensions we defined a collection $\{F_{ijkl}\}$ of vertex objects that are totally antisymmetric in their subscripts and that satisfy \begin{equation}F_{ijkl}+F_{jklm} +F_{klmi}+F_{lmij}+F_{mijk}=[ijklm]\label{Review4DMonodromy}\end{equation} for any choice of $i,j,k,l,m.$

The volume of any polytope is given by the sum over its vertices of these vertex objects.  This expression of the volume is unique, and any triangulation of the polytope can be recovered from this expression using (\ref{Review3DMonodromy}), (\ref{Review4DMonodromy}), and their higher-dimensional analogues.   Additionally, the expression of the volume of a polytope in terms of the vertex objects also encodes the geometry of all lower-dimensional boundary polytopes and readily gives their volumes as well~\cite{OldPaper}.

We note that equation (\ref{Review4DMonodromy}) is reminiscent of the formula 
\begin{equation}
 \partial [ijklm]=[ijkl]+[jklm]+[klmi]+[lmij]+[mijk] \label{NotTheSame}
\end{equation} 
given in \cite{NoteOnPolytopes}, describing the boundary $\partial [ijklm]$ of the simplex and encoding where the poles of $[ijklm]$ are.  Equation (\ref{Review4DMonodromy}) does the same, and is also a genuine equality between the volume of the simplex and objects that correspond to its vertices.  Thus the objects on the left of equation (\ref{Review4DMonodromy}) are fundamentally different than those on the right of equation (\ref{NotTheSame}).  Similar statements can be made about equations (\ref{Intro2DMonodromy}) and (\ref{Review3DMonodromy}) and the lower-dimensional analogues of equation (\ref{NotTheSame}).

\subsection{Applications to NMHV Amplitudes}

Quite surprisingly, the $n$-point NMHV tree-level superamplitude $M_{\rm NMHV}^n$ in $\mathcal{N}=4$ planar sYM can be written as the volume of a polytope in $\mathbb{CP}^{4*}$ \cite{NoteOnPolytopes}.  Indeed, $M_{\rm NMHV}^n$ can be represented as 
\begin{equation}
M_{\rm NMHV}^n= \sum_{i,j}^n[*i (i+1)j (j+1)]
\label{nPointAmp} 
\end{equation}
where the $\{Z_i^{\alpha}\}$ implicitly inside the five-brackets in the sum are $n$ points in $\mathbb{CP}^4$ encoding the external kinematics and $Z_*^{\alpha}$ is a reference vector in $\mathbb{CP}^4.$    The sum on $i,j$ is understood modulo $n.$   

For any given $n,$ $M_{\rm NMHV}^n$ has many different expressions depending on our choice of $Z^{\alpha}_*.$  For example, if we choose $Z^{\alpha}_*=Z^{\alpha}_1,$ then for $n=6$ we have \begin{equation}
M_{\rm NMHV}^6=[12345]+[12356]+[13456], \label{FirstForm}
\end{equation}
while if we choose $Z^{\alpha}_*=Z^{\alpha}_2,$ then we have\begin{equation}
M_{\rm NMHV}^6=[23456]+[23461]+[24561].\label{SecondForm}
\end{equation} 
Just as the relation (\ref{nonTrivialTriangles}) is not obvious at the level of Schouten identities on the $\langle ... \rangle$ brackets, the equivalence of the right hand sides of (\ref{FirstForm}) and (\ref{SecondForm}) is non-trivial.  These two representations of $M_{\rm NMHV}^6$ were initially found by performing two different BCFW shifts on the amplitude \cite{NoteOnPolytopes,SpuriousPoles,BCFW}.  The geometric interpretation is that they correspond to two different triangulations of the same underlying polytope.  As discussed in the introduction, their equality can also be understood by using a global residue theorem in an auxiliary Grassmannian \cite{MasonSkinner, DualityForSMatrix}.  Part of the utility of the vertex objects is to show that the right hand sides of (\ref{FirstForm}) and (\ref{SecondForm}) are equal directly---namely, they are identical when expressed in term of these objects.  By using equation (\ref{Review4DMonodromy}) on each simplex in either (\ref{FirstForm}) or (\ref{SecondForm}), we find 
\begin{equation}M_{\rm NMHV}^6=F_{1234}+F_{1245}+F_{1256}+F_{2345}+F_{2356}+F_{2361}+F_{3456}+F_{3461}+F_{4561}.\end{equation}
  The amplitude is therefore uniquely expressed in terms of the vertex objects.  From this expression and equation (\ref{Review4DMonodromy}), any triangulation of $M_{\rm NMHV}^6$ can be obtained. 

For general $n,$ we have 
\begin{align}
M_{\rm NMHV}^n&=\sum_{i,j}^n [*i(i+1)j(j+1)] \nonumber \\ 
&=\sum_{i,j}^n F_{*i(i+1)j}+F_{i(i+1)j(j+1)}+F_{(i+1)j(j+1)*}+F_{j(j+1)*i}+F_{(j+1)*i(i+1)} \nonumber  \\
&=\sum_{i,j}^n F_{i(i+1)j(j+1)},
\end{align}
where in the second equality we used equation (\ref{Review4DMonodromy}) and in the last equality we used the cyclicity of the sum and antisymmetry of the vertex objects to cancel in pairs any terms with $*$ as a subscript.  This shows manifestly that the amplitude is independent of $Z^{\alpha}_*$ and that the underlying polytope has vertices only where the four hyperplanes defined by $Z^{\alpha}_i,$ $Z^{\alpha}_{i+1},$ $Z^{\alpha}_j,$ and $Z^{\alpha}_{j+1}$ intersect.

We refer to Ref.~\cite{OldPaper} for further discussion of this vertex
formalism. In the next two sections we show that these vertex objects are
naturally defined as contour integrals of logarithms.

\section{Volumes and Logarithms}

In \cite{OldPaper} the vertex objects are defined as a particular sum of simplices.  Thus, in some sense, writing the volume of a polytope in terms of these objects may be viewed as simply choosing a particular triangulation.  However, we will now show that these objects are naturally defined in terms of contour integrals of logarithms, thus giving them an existence independent of simplices.  This further motivates the view that the vertex objects are basic building blocks for computing volumes of polytopes.

As mentioned in the introduction, our integrals differ from those discussed in Ref.~\cite{NoteOnPolytopes} in that the latter involve contours with boundaries on the underlying polytope.  Evaluating volumes in this way leads to the presence of spurious vertices (which correspond physically to spurious poles) associated to a particular triangulation.  For example, the vertex $\{1,2\}$ is a spurious vertex in the triangulation depicted in Figure \ref{DifferenceOfTriangles}, since it is not present in the underlying polytope but shows up in individual terms in the triangulation.  As we will see, the integrals we use have closed contours, so evaluating them corresponds to a straightforward application of Cauchy's residue theorem.  Moreover, they give rise to the vertex objects used in the vertex formalism discussed above, in which only the genuine (i.e., non-spurious) vertices of the polytope play a role.

\subsection{One Dimension}

As a warmup, we begin our discussion in one dimension.  Another way of writing the length $L$ of a line from $x_1$ to $x_2$ is as 
\begin{equation}
L=x_1-x_2=\int_{x_1\leq x \leq x_2}dx=\frac{1}{2\pi i}\int_{x_1\leq x \leq x_2}\ 2\pi i\ dx.
\label{IntegralLengthOfLine}
\end{equation}
By allowing the $x$ variables to be complex, we can define the complex logarithm function $\log\big(\frac{x-x_1}{x-x_2}\big)$ with its branch cut connecting the point $x_1$ to the point $x_2$ along the real axis.  We can then rewrite $2\pi i$ as $\text{Disc}\big(\log\big(\frac{x-x_1}{x-x_2}\big)\big)$---the discontinuity of the logarithm across its branch cut---giving
\begin{equation}
L=\frac{1}{2\pi i}\int_{x_1\leq x \leq x_2}\  \text{Disc}\Big(\log\Big(\frac{x-x_1}{x-x_2}\Big)\Big)\ \ dx.
\label{IntegralLengthOfLine2}
\end{equation}
Unwrapping the contour allows one to drop the ``Disc'' from the integrand and obtain
\begin{equation}
L=\frac{1}{2\pi i}\oint \  \log\Big(\frac{x-x_1}{x-x_2}\Big)\ \ dx.\label{IntegralLengthOfLine3}
\end{equation}
where the contour surrounds the cut.  Evaluating this explicitly (for example, by going around the pole at infinity) recovers $L=x_1-x_2,$ as expected.

Making the same definitions as in (\ref{1DDefinitions}) we can rewrite (\ref{IntegralLengthOfLine3}) as a contour integral in $\mathbb{CP}^{1*}$ as
\begin{equation}
L=\frac{1}{2\pi i}\oint \  \log\Big(\frac{Z_1\cdot X}{Z_2\cdot X}\Big) \frac{DX}{(P\cdot X)^2},
\label{IntegralLengthOfLine4}
\end{equation}
where $DX\equiv \varepsilon^{\alpha\beta}X_{\alpha}dX_{\beta}$ is the canonical volume form (of weight two) on $\mathbb{CP}^{1*}$ and $X_{\alpha}\equiv \begin{pmatrix} x\\1 \end{pmatrix}.$   By explicitly evaluating this integral we find 
\begin{equation}
L=\frac{\langle Z_1 Z_2\rangle}{\langle Z_1P\rangle\langle Z_2P\rangle},
\end{equation}
in agreement with equation (\ref{finalLengthOfLine}).  In this way, the length of a line is naturally represented as a contour integral of a logarithm. 

\subsection{Two Dimensions}
\label{TwoDims}

Motivated by the one-dimensional result, we consider the $\mathbb{CP}^{2*}$ integral \begin{equation}
A=\frac{1}{(2\pi i)^2}\oint \log\Big(\frac{Z_1\cdot X}{Z_2\cdot X}\Big) \log\Big(\frac{Z_3\cdot X}{Z_4\cdot X}\Big)\frac{DX}{(P\cdot X)^3},\label{TwoDimensionalLogs}
\end{equation}
where $DX\equiv\varepsilon^{\alpha\beta\gamma}X_{\alpha}dX_{\beta}dX_{\gamma}$ is the canonical volume form on $\mathbb{CP}^{2*}$ of weight three.  The contour is again defined by the integrand in a canonical way: first go around the branch cut of $\log\big(\frac{Z_3\cdot X}{Z_4\cdot X}\big)$ and then go around the branch cut of $\log\big(\frac{Z_1\cdot X}{Z_2\cdot X}\big)$.  This gives \begin{equation}A=[123]-[124],\end{equation} which is precisely the area of the quadrilateral given in equation (\ref{InformalQuadrilateralArea}).  If we swap $Z^{\alpha}_3,$ and $Z^{\alpha}_4$ with $Z^{\alpha}_1$ and $Z^{\alpha}_2$ in equation (\ref{TwoDimensionalLogs}) and pick up a minus sign from the change in orientation of the contour, one readily sees that  \begin{equation}A=-([341]-[342]),\end{equation} thus proving the identity $[123]-[123]=[431]-[432]$ that we obtained in section \ref{GeneralPolys}.  This identity is now made manifest by the integrand of (\ref{TwoDimensionalLogs}).

We have expressed a two-dimensional area as a closed contour integral whose contour specification comes naturally with the integrand itself.  The objects whose area we compute in this way are quadrilaterals, defined by four lines.  Before describing how the vertex objects are obtained from these kinds of integrals, we quickly discuss how we can use these integrals to compute the volume of three- and $D$-dimensional ``quadrilaterals,'' or hypercubes.

\subsection{Higher Dimensions}
\begin{figure}[h!]
\centering
    \includegraphics[scale=.7]{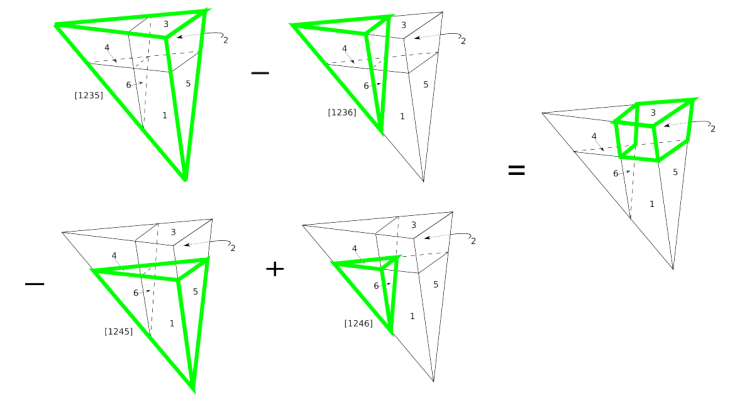}
        \caption{A triangulation of the cube using four simplices.}\label{TriangulationOfCube}
\end{figure}

Consider the following contour integral in $\mathbb{CP}^{3*}:$
\begin{equation}
V=\frac{1}{(2\pi i)^3}\oint \log\Big(\frac{Z_1\cdot X}{Z_2\cdot X}\Big)\log\Big(\frac{Z_3\cdot X}{Z_4\cdot X}\Big)\log\Big(\frac{Z_5\cdot X}{Z_6\cdot X}\Big)\frac{DX}{(P\cdot X)^4},\label{3DCube}
\end{equation}
 where $DX\equiv\varepsilon^{\alpha\beta\gamma\delta}X_{\alpha}dX_{\beta}dX_{\gamma}dX_{\delta}.$   The contour is a three-torus $(S^1)^3$ that goes around the branch cut of each logarithm.  We find that \begin{equation}V=[1235]-[1236]-[1245]+[1246].\label{VolumeOfCube}\end{equation}  This corresponds to the volume of a three-dimensional ``cube,'' where we simply mean a polytope bounded by 3 pairs of faces.  One way to see that equation (\ref{VolumeOfCube}) is triangulating a ``cube'' with faces 1 and 2 opposite each other, 3 and 4 opposite each other, and 5 and 6 opposite each other is by examining Figure \ref{TriangulationOfCube}, which shows the superposition of the four simplices in (\ref{VolumeOfCube}) leaving the volume of a ``cube.''

\begin{figure}[h!]
\centering
    \includegraphics[scale=.6]{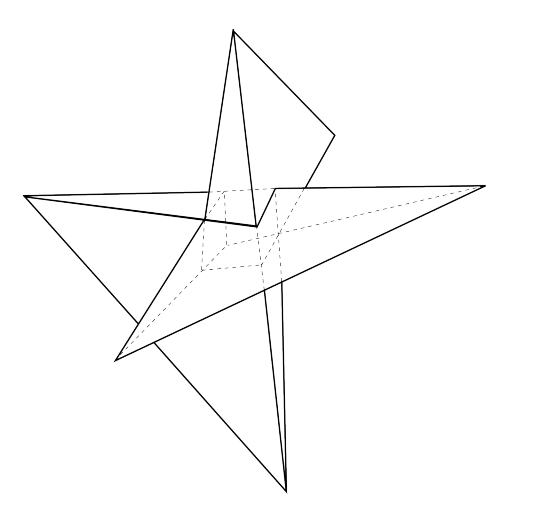}
        \caption{The three-dimensional analogue of Figure \ref{DifferenceOfTriangles}, showing the three possible ways of forming a triangulation analogous to that shown in Figure \ref{TriangulationOfCube}.}\label{DifferentCubeTriangulations}
\end{figure}

As in the two-dimensional case, there is more than one expression for the volume of this cube.  Namely, just as we could get two different expressions for the area of a quadrilateral by viewing it as the difference between two different pairs of triangles, we can get three expressions for the volume of the cube as a superposition of four simplices.  In particular, we also have \begin{equation}V=-([3415]-[3416]-[3425]-[3426]) \ \ \text{and}\ \ \ V=-([5631]-[5632]-[5641]-[5642]),\end{equation} which correspond to the different ways of decomposing the cube analogously to Figure \ref{TriangulationOfCube} corresponding to Figure \ref{DifferentCubeTriangulations}.  Figure \ref{DifferentCubeTriangulations} is the three-dimensional analog of Figure \ref{DifferenceOfTriangles}.  As in the two-dimensional case, these identities are manifest from the integrand in (\ref{3DCube}) by swapping, for example,  $Z_1^{\alpha}$ and $Z_2^{\alpha}$ with $Z_3^{\alpha}$ and $Z_4^{\alpha},$ or with $Z_5^{\alpha}$ and $Z_6^{\alpha},$ and picking up a minus sign from the change in orientation of the contour.

For completeness we write down the contour integral that gives the volume of a $D$-dimensional ``hypercube'' bounded by $2D$ faces in ``pairs.''  With $Z_1^{\alpha}, ..., Z_{2D}^{\alpha}$ defining the $2D$ faces, we have a generalization of the lower-dimensional cases:
\begin{equation}
V=\frac{1}{(2\pi i)^D}\oint \log\Big(\frac{Z_1\cdot X}{Z_2\cdot X}\Big)...\log\Big(\frac{Z_{2D-1}\cdot X}{Z_{2D}\cdot X}\Big)\frac{DX}{(P\cdot X)^{D+1}},
\end{equation}
where $DX$ is the natural generalization of the lower-dimensional volume forms and the contour goes around the branch cut of each logarithm.  We now turn our attention to using these types of objects to obtain the vertex objects and thus to compute the volumes of general polytopes.

\section{Vertex Objects from Logarithms}

We motivate the vertex objects by first seeing how to recover the volume of a simplex from integrals of logarithms.  We define \begin{equation} T_{12}\equiv\frac{1}{(2\pi i)^2}\oint_{\gamma_{12}} \log\Big(\frac{Z_1\cdot X}{Z_2\cdot X}\Big)\log\Big(\frac{Z_{3}\cdot X}{Q\cdot X}\Big)\frac{DX}{(P\cdot X)^{3}}=[123]-[12Q],\end{equation} where $\gamma_{12}$ is the same contour that we have described before, only now we are making it explicit.  We have also introduced a fixed reference vector $Q^{\alpha},$ defining a reference line in $\mathbb{CP}^{2*}.$  Cyclicly permuting 1, 2, and 3, we define \begin{equation} T_{23}\equiv\frac{1}{(2\pi i)^2}\oint_{\gamma_{23}} \log\Big(\frac{Z_2\cdot X}{Z_3\cdot X}\Big)\log\Big(\frac{Z_{1}\cdot X}{Q\cdot X}\Big)\frac{DX}{(P\cdot X)^{3}}=[231]-[23Q],\end{equation} as well as \begin{equation} T_{31}\equiv\frac{1}{(2\pi i)^2}\oint_{\gamma_{31}} \log\Big(\frac{Z_3\cdot X}{Z_1\cdot X}\Big)\log\Big(\frac{Z_{2}\cdot X}{Q\cdot X}\Big)\frac{DX}{(P\cdot X)^{3}}=[312]-[31Q].\end{equation}  It is important to note that $\gamma_{12},$ $\gamma_{23},$ and $\gamma_{31}$ are all different contours, each being the contour defined by the integrand of the corresponding integral.  Performing these integrations, we find that \begin{equation}T_{12}+T_{23}+T_{31}=2[123].\label{SumOfStars}\end{equation}

The dependence on $Q^{\alpha},$ while present in each $T_{ij},$ drops out of this sum and we are left with twice the volume of a single simplex.  In the next subsection we use integrals similar to those defining the $T_{ij}$'s to define the vertex objects.

\subsection{Two-Dimensional Vertex Objects}

Suppose that we have $N$ points $\{Z_i^{\alpha}\}_{1\leq i\leq N}$ in $\mathbb{CP}^2,$ each defining a line in $\mathbb{CP}^{2*}.$  We define the following collection of $\binom{N}{2}$ integrals:
\begin{equation}
F_{ij}\equiv\frac{1}{2}\frac{1}{(2\pi i)^2}\oint_{\gamma_{ij}} \log\Big(\frac{Z_i\cdot X}{Z_j\cdot X}\Big)\Big(\sum_{k\neq i,j}\log\Big(\frac{Z_{k}\cdot X}{Q\cdot X}\Big)\Big)\frac{DX}{(P\cdot X)^{3}}\equiv \frac{1}{(2 \pi i)^2}\oint_{\gamma_{ij}} f_{ij}(X)\frac{DX}{(P\cdot X)^{3}}, \label{DefinitionOfF2D}
\end{equation}
where the contour $\gamma_{ij}$ first goes around all of the branch cuts from $Z_k\cdot X=0$ to $Q\cdot X=0$ and then around the branch cut from $Z_i\cdot X=0$ to $Z_j\cdot X=0.$  The factor of $\frac{1}{2}$ is conventional.  Integrating this gives \begin{equation}
F_{ij}=\frac{1}{2}\sum_{k\neq i,j}^N([ijk]-[ijQ])=\frac{1}{2}\Big(\sum_{k\neq i,j}^N[ijk]\Big)-\frac{1}{2}(N-2)[ijQ]\label{FInTermsOfSimplices}.
\end{equation} These are (up to a factor of 2) the vertex objects of Ref.~\cite{OldPaper} and so in particular we have, for any $i,j,k\in \{1, ..., N\},$ that  \begin{equation}F_{ij}+F_{jk}+F_{ki}=[ijk].  \label{2DMonodromy}\end{equation} Each individual vertex object depends on $Q^{\alpha}$ as well as all $N$ of the $Z_i^{\alpha}$'s, but the dependence on $Q^{\alpha}$ and all other $Z_l^{\alpha}$'s (i.e., for $l\neq i,j,k$) drops out in the above sum. 

In equation (\ref{DefinitionOfF2D}) we wrote $F_{ij}$ as an integral over a function $f_{ij}(X)$ on the dual space.  We readily see that $f_{ij}=-f_{ji},$ and can also show that for any choice of $i,$ $j,$ and $k,$ \begin{equation}f_{ij}+f_{jk}+f_{ki}=0. \label{2DCohomology} \end{equation}  The antisymmetry of the $f_{ij}$'s as well as property (\ref{2DCohomology}) imply that the collection of functions $\{f_{ij}\}$ form a representative of a $\check{C}$ech cohomology class on a subspace of $\mathbb{CP}^{2*}.$

In twistor theory, $\check{C}$ech cohomology is a natural setting in which to discuss the Penrose transform, which takes a cohomology class on (a subspace of) twistor space to a finite-normed on-shell field configuration on space-time~\cite{Huggett}.  The appearance of $\check{C}$ech cohomology here is of a different nature, and the role it is playing in this discussion is still unclear. For the remainder of this note we will not explore this issue.  Instead, we simply note this curious connection to cohomology, as it may be important for generalizing these ideas to the N$^k$MHV amplituhedron with $k>1.$  For now, we simply move on to describing how to construct the higher-dimensional vertex objects in terms of integrals of logarithms.

\subsection{Higher-Dimensional Vertex Objects}

Analogous vertex objects can be defined in any dimension.  Namely, in $D$ dimensions there exist objects $F_{i_1...i_D}$ such that for any choice of $D+1$ hyperplanes defined by $\{Z_{i_k}\}_{1\leq k \leq D+1},$ one has the identity \begin{equation}F_{i_1i_2...i_D}+(-1)^DF_{i_2i_3...i_{D+1}}+F_{i_3i_4...i_1}+...+(-1)^DF_{i_{D+1}i_1...i_{D-1}}=[i_1i_2...i_{D+1}]. \end{equation} 
Given any polytope in $\mathbb{CP}^D,$ one obtains its volume by summing the vertex objects over the vertices of the polytope.  In particular, any vertex of the polytope is defined (as reviewed in section 2) by the intersection of $D$ hyperplanes corresponding to $Z_{i_1},...,Z_{i_D},$ and for this vertex one simply includes an $F_{i_1...i_D}.$  The precise definition of higher-dimensional polytopes in $\mathbb{CP}^D$ is described in Ref.~\cite{OldPaper}, as is the precise way of summing the vertex objects over the vertices.  In this subsection, we will see how these higher-dimensional vertex objects arise as contour integrals of logarithms.   We will explicitly show this only for dimensions three and four.

\subsubsection{Three Dimensions}

Let $\{Z_i^{\alpha}\}_{1\leq i \leq N}$ be $N$ points in $\mathbb{CP}^3$ defining $N$ planes in the dual $\mathbb{CP}^{3*}.$  Motivated by the two-dimensional case, we define \begin{equation}F_{ij;k}\equiv\frac{1}{(2\pi i)^3}\oint_{\gamma_{ij;k}}\log\Big(\frac{Z_i\cdot X}{Z_j \cdot X}\Big)\log\Big(\frac{Z_k\cdot X}{Q_2\cdot X}\Big)\sum_{l\neq i,j,k}\log\Big(\frac{Z_l\cdot X}{Q_1\cdot X}\Big)\frac{DX}{(P\cdot X)^4}, \end{equation} where $Q_1^{\alpha}$ and $Q_2^{\alpha}$ are fixed reference points in $\mathbb{CP}^3$ defining fixed reference planes in $\mathbb{CP}^{3*}.$  The contour $\gamma_{ij;k}$ is an $(S^1)^3$ contour going around the branch cuts of the logarithms in the natural way.  Antisymmetrizing over $i,$ $j,$ and $k,$ and noting that each $F_{ij;k}$ is antisymmetric in its first two indices, we then define \begin{equation}F_{ijk}\equiv \frac{1}{2\cdot3!}F_{[ij;k]}=\frac{1}{3!}(F_{ij;k}+F_{jk;i}+F_{ki;j}).\end{equation}  
Each $F_{ijk}$ dependends on $Q^{\alpha}_1$ and all $N$ of the $Z_i^{\alpha}$'s, although it turns out that it is independent of $Q^{\alpha}_2.$  We also show that for any chocie of $i,j,k,l\in \{1, ..., N\},$ one has \begin{equation}F_{ijk}-F_{jkl}+F_{kli}-F_{lij}=[ijkl],\label{3DMonodromy}\end{equation} where $[ijkl]$ is the volume of the three-simplex bounded by the four faces defined by $Z^{\alpha}_i,$ $Z^{\alpha}_j,$ $Z^{\alpha}_k,$ and $Z^{\alpha}_l.$  The dependence on $Q_1^{\alpha}$ and all other $Z^{\alpha}_m$'s drops out in this sum.  

\subsubsection{Four Dimensions}

 The definition of the four-dimensional vertex objects is similar.  Let $\{Z_i^{\alpha}\}_{1\leq i \leq N}$ be $N$ points in $\mathbb{CP}^4$ defining $N$ hyperplanes in the dual $\mathbb{CP}^{4*}.$  Define \begin{equation}F_{ij;k;l}=\frac{1}{(2\pi i)^4}\oint_{\gamma_{ij;k;l}}\log\Big(\frac{Z_i\cdot X}{Z_j \cdot X}\Big)\log\Big(\frac{Z_k\cdot X}{Q_3\cdot X}\Big)\log\Big(\frac{Z_l\cdot X}{Q_2\cdot X}\Big)\sum_{m\neq i,j,k,l}\log\Big(\frac{Z_m\cdot X}{Q_1\cdot X}\Big)\frac{DX}{(P\cdot X)^5}, \end{equation} where $Q^{\alpha}_1,$ $Q^{\alpha}_2,$ and $Q^{\alpha}_3$ are fixed reference points in $\mathbb{CP}^4$ defining reference hyperplanes in $\mathbb{CP}^{4*}.$  The contour $\gamma_{ij;k;l}$ is an $(S^1)^4$ contour going around the branch cuts of the logarithms in the natural way.  We define \begin{align}F_{ijkl}\equiv \frac{1}{2\cdot 4!}F_{[ij;k;l]}=\frac{1}{4!}(&F_{ij;k;l}-F_{ij;l;k}+F_{ik;l;j}-F_{ik;j;l}+F_{il;j;k}-F_{il;k;j}\\ \nonumber &+F_{jk;i;l}-F_{jk;l;i}+F_{jl;k;i}-F_{jl;i;k}+F_{kl;i;j}-F_{kl;j;i}).\end{align} Each individual $F_{ijkl}$ is independent of $Q_2^{\alpha}$ and $Q_3^{\alpha},$ though it is dependent on $Q_1^{\alpha}$ and all $N$ of the $Z_i^{\alpha}$'s.  For any choice of $i,$$j,$$k,$$l,$ and $m,$ we have \begin{equation}F_{ijkl}+F_{jklm}+F_{klmi}+F_{lmij}+F_{mijk}=[ijklm],  \label{4DMonodromy}\end{equation} where $[ijklm]$ is the volume of a four-simplex bounded by the five faces defined by $Z^{\alpha}_{i},$ $Z^{\alpha}_{j},$ $Z^{\alpha}_{k},$ $Z^{\alpha}_{l},$ and $Z^{\alpha}_{m}.$   Again, the dependence on $Q_1^{\alpha}$ and all other $Z^{\alpha}_n$'s drops out in this sum.  The three- and four-dimensional vertex objects introduced in this section are equal (up to a factor of $2\cdot 3!$ and $2\cdot 4!,$ respectively) to the vertex objects introduced in Ref.~\cite{OldPaper}. 
 
 \section{Conclusion and Outlook}

In this paper we showed that volumes of general polytopes can be computed using contour integrals of logarithms directly in the space in which the polytopes live.  The contours of these integrals are canonically specified by the integrands themselves, and the organizing principle for combining these integrals comes directly from the geometry of the polytope---the intersections of its faces---and thus does not rely on any particular triangulation.  We also found a surprising connection between the integrands of the two-dimensional vertex objects and $\check{C}$ech cohomology.  It would interesting to further explore this connection.

The vertex objects that we have defined are useful for computing NMHV tree-level amplitudes in the planar limit of $\mathcal{N}=4$ super-Yang--Mills, and we have seen logarithms appear naturally.  It would be interesting to see how these ideas might generalize to loop level.  Additionally, since our discussion has been limited to tree-level amplitudes, these results readily apply at tree level to Yang--Mills theories with less (and no) supersymmetry.  It would therefore be interesting to see if similar ideas can be used for less supersymmetric theories beyond tree level.  Taking the planar limit appears to be crucial in this discussion, as momentum (super-)twistors play a fundamental role and these cease to exist in non-planar theories.  Nonetheless, it is worth exploring if and to what extent this discussion can be extended to the non-planar sector of the theory.

The vertex objects we defined can be used to obtain identities amongst sums of simplices, and these identities can therefore now be viewed as being obtained from contour integrals of logarithms directly in the space containing the polytope.  This differs dramatically from the Grassmannian picture discussed in the introduction.  Understanding the relation between these two approaches will help extend the method introduced in this note to N$^k$MHV tree amplitudes for $k>1,$ since the Grassmannian picture is already well-understood for these more complicated cases.  Expressing volumes in terms of the vertex objects naturally encodes the geometry of the underlying polytope.  If the analogous objects can be found for the $k>1$ cases, likely by first making a connection to the Grassmannian picture, then this should shed light on the geometry of the dual amplituhedron directly, without a need for any auxiliary spaces.

\acknowledgments

I thank Zvi Bern for his help.  I would also like to thank Andrew Hodges, and I am grateful to Julio Parra Martinez and Enrico Herrmann for many interesting discussions.  I am grateful to the Mani L. Bhaumik Institute for Theoretical Physics for its support.

\end{document}